\documentclass[conference]{IEEEtran}
\IEEEoverridecommandlockouts
\usepackage{cite}
\usepackage{amsmath,amssymb,amsfonts}
\usepackage{algorithmic}
\usepackage{graphicx}
\usepackage{textcomp}
\usepackage[dvipsnames]{xcolor}
\usepackage{booktabs}
\usepackage{multirow}
\usepackage{mathrsfs}
\usepackage{fancyhdr}

\usepackage{bbold}
\usepackage{booktabs}
\usepackage{caption}
\usepackage{enumitem}
\usepackage{float}
\usepackage{makecell}
\usepackage{multicol}
\usepackage{pgfplots,lipsum}
\usepackage{subcaption}
\usepackage{tikz,tikzscale}

\def\BibTeX{{\rm B\kern-.05em{\sc i\kern-.025em b}\kern-.08em
    T\kern-.1667em\lower.7ex\hbox{E}\kern-.125emX}}

\begin{document}

\title{\textit{FedABC}: Attention-Based Client Selection for Federated Learning with Long-Term View \\
}

\author{
\IEEEauthorblockN{Wenxuan Ye\IEEEauthorrefmark{1}\IEEEauthorrefmark{2},
Xueli An\IEEEauthorrefmark{1}, 
Junfan Wang\IEEEauthorrefmark{3},
Xueqiang Yan\IEEEauthorrefmark{3},
Georg Carle\IEEEauthorrefmark{2}}
\IEEEauthorblockA{
\textit{\IEEEauthorrefmark{1} Advanced Wireless Technology Laboratory, Munich Research Center, Huawei Technologies Duesseldorf GmbH}\\
\textit{\IEEEauthorrefmark{2} TUM School of Computation, Information and Technology, Technical University of Munich}\\
\textit{\IEEEauthorrefmark{3} Wireless Technology Lab, 2012 Laboratories, Huawei Technologies Co., Ltd}\\
}
wenxuan.ye@tum.de, \{xueli.an, wangjunfan3, yanxueqiang1\}@huawei.com, carle@net.in.tum.de
}

\maketitle


\begin{abstract}
Native AI support is a key objective in the evolution of 6G networks, with Federated Learning (FL) emerging as a promising paradigm. 
FL allows decentralized clients to collaboratively train an AI model without directly sharing their data, preserving privacy. 
Clients train local models on private data and share model updates, which a central server aggregates to refine the global model and redistribute it for the next iteration.
However, client data heterogeneity slows convergence and reduces model accuracy, and frequent client participation imposes communication and computational burdens.
To address these challenges, we propose \textit{FedABC}, an innovative client selection algorithm designed to take a long-term view in managing data heterogeneity and optimizing client participation.
Inspired by attention mechanisms, \textit{FedABC} prioritizes informative clients by evaluating both model similarity and each model's unique contributions to the global model.
Moreover, considering the evolving demands of the global model, we formulate an optimization problem to guide \textit{FedABC} throughout the training process.
Following the ``later-is-better" principle, \textit{FedABC} adaptively adjusts the client selection threshold, encouraging greater participation in later training stages.
Extensive simulations on CIFAR-10 demonstrate that \textit{FedABC} significantly outperforms existing approaches in model accuracy and client participation efficiency, achieving comparable performance with 32\% fewer clients than the classical FL algorithm \textit{FedAvg}, and 3.5\% higher accuracy with 2\% fewer clients than the state-of-the-art.
This work marks a step toward deploying FL in heterogeneous, resource-constrained environments, thereby supporting native AI capabilities in 6G networks.
\end{abstract}


\section{INTRODUCTION}
AI is recognized as a pivotal force in evolving 6G network architectures. 
Beyond utilizing AI in optimizing intelligent communication systems, 6G aims to achieve native AI support, effectively elevating networks from mere channels of transmitting collected data \cite{Tong2021}. 
Federated Learning (FL) stands out as a promising learning paradigm, which enables communication-effective and privacy-preserving data analysis across multiple decentralized clients \cite{McMahan2017, Ye2023}. 
Generally, after clients perform local model training on their private data and share model updates, the server aggregates these updates to refine the global model, which is redistributed to clients for the next training iteration.
It holds diverse potentials across vertical industries, such as smart cities and healthcare, by facilitating extensive data analysis without compromising privacy \cite{Kairouz2021}. 
Recent progress in decentralized computing and communication capabilities of mobile networks further bolsters FL implementations \cite{Ye2022}.

Despite its potential, \textit{the heterogeneous nature of client local data} presents significant challenges. 
This heterogeneity may arise from variations in client behaviors, geographic locations, and device-specific data characteristics.
Such heterogeneity leads to inconsistent and even conflicting client updates, extending convergence and degrading the precision of the global model \cite{Kairouz2021}.
To mitigate the impact of data heterogeneity, research has explored various methods, such as regularization \cite{Li2020}, data clustering methods \cite{Sattler2020}, graph-based analysis \cite{Qian2022} and sharing small subsets of data globally to align distributions \cite{Zhao2018}. 
Although these approaches have shown promise, they often overlook the significant \textit{computational and communication demands} placed on resource-limited clients.
Together, these factors highlight the inherent complexities and operational challenges of effectively deploying FL in real-world scenarios.

Tackling the challenge of data heterogeneity while reducing the burden on client resources requires novel and adaptive strategies.
A promising solution involves refining client selection methods, as carefully choosing clients to participate in each iteration round can help mitigate the adverse effects of data heterogeneity \cite{Kairouz2021, Cho2022}.
Efforts include AUCTION algorithms with analysis of the data distribution \cite{Deng2021}, methods prioritizing clients according to the time consumed in local computations \cite{Nishio2019}, or cluster methods to group clients with similar data distributions \cite{Sattler2020}. 
Despite these advancements, many existing algorithms either treat client models as isolated units, or only focus on data similarity.
Furthermore, many approaches rely on static criteria, lacking the flexibility to adapt to the evolving demands of the global model and changing network conditions in communication networks.

To adaptively manage data heterogeneity and optimize client participation throughout the entire FL training process, we propose an innovative client selection method, \textit{FedABC}. 
At its core, \textit{FedABC} draws inspiration from attention mechanisms \cite{Vaswani2017, Kim2017}, which assign different weights to parts of the input based on their relevance to the task, enabling the model to focus on critical information. 
Similarly, \textit{FedABC} dynamically prioritizes client models that contribute the most value to the global model, by leveraging two key aspects: \textit{client data similarity} and \textit{unique client contributions}. 
Data similarity is evaluated by aligning client model predictions, assuming that similar prediction patterns indicate related data distributions among clients. 
Unique contributions are quantified by the server model's loss on each client's local dataset \cite{Cho2022}, with higher local loss suggesting valuable, under-represented information not yet captured by the global model.
By integrating both aspects, \textit{FedABC} effectively selects the most representative and valuable clients, enhancing the global model's performance and accelerating convergence.

To guide \textit{FedABC} in the long-term view, we formulate an optimization problem to manage client participation effectively throughout the training.
Building on the ``later-is-better" insights from \cite{Xu2020} that highlight the greater impact of client involvement in later training stages, we gradually encourage more client participation over time.
By strategically distributing client participation among the whole FL training process, this approach efficiently utilizes resources and enhances learning.

Our contributions are highlighted as follows:
\begin{itemize}
    \item {To address data heterogeneity and optimize client participation, we propose a novel client selection algorithm, drawing inspiration from attention mechanisms.
    This approach identifies underrepresented information and prioritizes clients accordingly by utilizing client data similarity and distinctive model contributions.
    }
    \item {To accommodate the varying demands of the FL system, our algorithm incorporates a long-term, adaptive selection strategy. 
    Following the ``later-is-better" principle, it progressively encourages client diversity in later training stages, enhancing the model’s exposure to a wide range of data distributions over time. 
    }
    \item {Extensive empirical evaluations demonstrate that our algorithm significantly outperforms existing methods, achieving higher model accuracy and meanwhile fewer client participation. 
    Additionally, we incorporate a comprehensive cost analysis to further evaluate the practicality of our approach.
    }
\end{itemize}

\section{PRELIMINARIES}
\subsection{Federated Learning (FL)}
FL is a decentralized learning approach that enables a server to learn from clients without exposing clients' raw data.
Each client $k$ trains a local model using its own private data $\mathbb{D}_{k}=\{(x^i_k, y^i_k)\}_i$, by minimizing the following objective function:
\begin{equation}
\theta_{k} = \arg \min\nolimits_{\theta} \mathbb{E}_{(x^i_k, y^i_k) \sim \mathbb{D}_{k}}[\mathcal{F}(y^i_k, f(\theta,x^i_k))]
\end{equation}
where $\mathcal{F}(\cdot)$ represents the chosen loss function; $f(\theta,x^i_k)$ represents the model’s output for input $x^i_k$ under parameters $\theta$; 
$\mathbb{E}[\cdot]$ denotes the expectation function.
Once training is complete, the client shares its updated model parameters $\theta_{k}$ with the server.

The server aggregates these parameters from participating clients to update the global model $\theta_{s}$. 
This aggregation process $\mathcal{A}(\cdot)$ is mathematically represented as follows:
\begin{equation} 
\label{eq:a}
\theta_s = \mathcal{A}(\{\theta_k\}_k) = \frac{1}{\sum_{k}w_k} \sum_{k} w_k \theta_{k} 
\end{equation}
where $w_k$ represents the weight assigned to each client $k$, with $w_k \geq 0$. 

The weights can be determined based on various factors, including the volume of data, the quality of the data, or the performance of the local model. 
This weighted aggregation helps to optimize the overall learning process by emphasizing contributions from more reliable or informative sources. 

\subsection{Attention Mechanism}
The attention mechanism \cite{Vaswani2017} is a pivotal innovation in neural networks, designed to address the limitations of sequence length in modeling dependencies. 
It dynamically focuses on the most relevant segments of the input data, allowing for the selective prioritization of information \cite{Kim2017}.

The attention mechanism comprises compatibility scores and values, built from queries $Q$ and keys $K$ of dimension $d_{k}$, and values $V$ of dimension $d_v$. 
Queries seek relevant information for specific parts of the model's inputs, and keys enable retrieval by matching these queries.
Compatibility scores are calculated by comparing queries and keys to determine relevance. 
Values hold the actual input information.
The retrieval process effectively uses these compatibility scores to prioritize and select relevant information from the values, facilitating focused model inference.

Mathematically, the process of the attention mechanism can be described as a weighted sum of the values.
This begins with the computation of a compatibility score by taking the dot product of the query with keys, scaled by $\sqrt{d_k}$ to stabilize the training process. 
The weights are then normalized using a softmax function, denoted as $\sigma(\cdot)$. 
The complete attention operation is expressed as:
\begin{equation}
\text{Attention(Q, K, V)} = \sigma(\frac{QK^T}{\sqrt{d_k}})V 
\end{equation}
This formulation effectively encapsulates how attention mechanisms leverage the interplay between queries, keys, and values to focus on the most informative parts of the input.

\section{SYSTEM MODEL}

The system consists of a central server and $K$ clients, each client $k$ having its own private dataset $\mathbb{D}_{k}$.
The server holds a test dataset $\mathbb{D}_{s, t}$, and a small portion of the training dataset $\mathbb{D}_{s, u}$, which is used for client selection and may be unlabeled. 
The FL training process involves iterative exchanges of model parameters between the clients and the server, with each complete exchange constituting one global round.

In the initial global round ($t=0$), the server mandates participation from all the clients and distributes the model architecture, which includes specifications such as layer details, model types, and neuron counts. 
Each client initializes its local model $\theta_{k}^{0}$, and sends it to the server, which stores the model as $\theta_{k}^{*}$. 
In subsequent global rounds, when a client $k$ is selected and updates its local model, the server replaces $\theta_{k}^{*}$ with the updated version.
 
Aligning with the FL protocol designed by \cite{Bonawitz2019}, each subsequent global round of training is systematically divided into three main steps: configuration, selection, and report. 
In the global round $t$, during the \textit{configuration} step, the server distributes the current global model parameters $\theta_s^{t-1}$ to clients. 
During the \textit{selection} step, the server employs a specifically designed algorithm for client selection, which we detail in Sec.~\ref{sec:design}. 
Lastly, in the \textit{report} step, the selected clients perform local computations to update their model parameters based on their datasets and send these updates $\theta_{k}^{t}$ back to the server. 
The server then aggregates these local updates to enhance the global model, marking the completion of this global round.

In the classical FL algorithm, \textit{FedAvg}, all clients are selected in each global round, and their local models are averaged to generate the global model \cite{McMahan2017}.
In contrast, our method employs selective client participation based on their contributions, updating the global model through weighted aggregation of the selected local models. 

\noindent\textbf{Application scenario:}
In this work, we focus on classification tasks with $N$ classes, aiming to optimize a global model across decentralized clients with heterogeneous data distributions. 
For local training, we use the cross-entropy (CE) loss as the objective function, given its effectiveness and widespread use in classification, i.e., $\mathcal{F}(\cdot)$ defined in Eq. (1) adopts CE loss $\text{CE}(\cdot)$. 
The loss function for the local training of client $k$ is defined as:
\begin{equation}
\label{eq:l}
\mathcal{L}(\mathbb{D}_{k}, \theta_k) = \mathbb{E}_{(x^i_k, y^i_k) \sim \mathbb{D}_{k}}[\text{CE}(y^i_k, f(\theta_k, x^i_k))]
\end{equation}

\section{DESIGN OF FedABC}
\label{sec:design}

\begin{figure}[t!]
    \begin{center}
    \includegraphics[width=0.8\columnwidth]{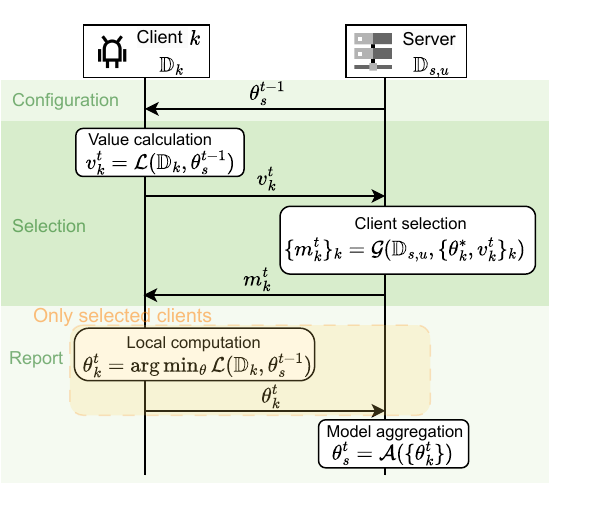}
    \caption{\textit{FedABC} procedure, which introduces a novel selection mechanism.
    In global round $t$, after the \textit{configuration} step, each client $k$ generates the value $v^t_k$ based on the global model $\theta_{s}^{t-1}$ and its data $\mathbb{D}_{k}$, and shares $v^t_k$ with the server.
    Then, the server selects clients by analyzing their values $v^t_k$ and their latest local models $\theta^{*}_{k}$ using its dataset $\mathbb{D}_{s, u}$, and distributes the selection indicators to the clients.
    In the \textit{report} step, selected clients update their local models, which the server aggregates to refine the global model.
    }
    \label{fig:arch}
    \end{center}
\end{figure}

This section is dedicated to introducing our proposed \textit{FedABC} algorithm, with the procedure illustrated in Fig.~\ref{fig:arch}.

\subsection{Problem Formulation}
Prior work \cite{Yang2020} has demonstrated that selecting more clients per global round can enhance FL performance, while resource constraints often hinder complete client participation.
Moreover, each client may not consistently contribute equally to global model performance. 
Inspired by \cite{Xu2020, Bi2025}, we introduce the following metric to represent FL performance at global round $t$:
\begin{equation}
    \eta^t \sum_{k=0}^{K-1} m_{k}^{t} \: \mathcal{S}(\theta_{k}^{t})
\end{equation}
Here, $\eta^t \in (0, 1]$ represents a temporal weighting factor for global round $t$;
$m_{k}^{t} \in \{0, 1\}$ is a binary decision variable for client $k$ in the global round $t$, where $m_{k}^{t} = 1$ if client $k$ is selected, $m_{k}^{t} = 0$ otherwise;
The function $\mathcal{S}(\theta_{k}^{t})$ measures the attention score of client model $\theta_{k}^{t}$ to the global model.
Defining $\mathcal{S}$ appropriately is crucial for optimizing client selection and ensuring efficient learning. 
In this work, we introduce a novel score function, which will be elaborated on in Sec. \ref{sec:score}.

Following the ``later-is-better" principle from \cite{Xu2020}, we emphasize the increasing importance of later training stages for model convergence. Thus, we increase $\eta^t$ with $t$, reflecting the growing significance of these rounds in the training process.

In FL scenarios within communication networks, clients often operate under resource constraints, and training and sharing local models can impose substantial computational and communication demands on each client.
Thus, minimizing client participation is also a key objective to reduce resource consumption. 
The corresponding optimization target is given by: 
\begin{equation} 
 \min \sum_{t=0}^{T-1} \sum_{k=0}^{K-1} m_{k}^{t} 
\end{equation}

Our objective is to maximize the cumulative attention scores of the selected clients over all the global rounds while minimizing client participation.
We formulate this as a single-objective optimization problem by introducing a regularization parameter $\lambda > 0$ to balance these two aspects:
\begin{equation}
\label{eq:op}
    \max \: \sum_{t=0}^{T-1} \left( \eta^t \sum_{k=0}^{K-1} m_{k}^{t} \: \mathcal{S}(\theta_{k}^{t})\right) - \lambda \sum_{t=0}^{T-1} \sum_{k=0}^{K-1} m_{k}^{t}
\end{equation}

To solve this optimization problem, we reformulate Eq.(~\ref{eq:op}) into an online optimization framework where decisions are made at each global round based on current and past information. 
Specifically, at each global round $t$, we solve the following optimization problem:
\begin{equation}
\label{eq:optt}
    \max \: \eta^t \sum_{k=0}^{K-1} m_{k}^{t} \: \mathcal{S}(\theta_{k}^{t}) - \lambda \sum_{k=0}^{K-1} m_{k}^{t}
\end{equation}
By dynamically adjusting client participation and resource allocation using current data, we optimize the FL process in a scalable and practical manner.

\subsection{Attention Score Algorithm}
\label{sec:score}
The attention score algorithm, denoted as $\mathcal{S}(\cdot)$, utilizes the latest local model $\theta_k^{*}$ and the value $v_k^{t}$ for each client, and the server dataset $\mathbb{D}_{s,u}$.
Aimed at identifying and prioritizing the most informative client model, we build a novel value assignment method on attention mechanisms.
As outlined in the preliminaries section, the attention mechanism is comprised of two pivotal components: compatibility scores and values. 

\noindent\textbf{Compatibility scores:}
These scores measure the informational similarity between client data. 
To evaluate data distribution similarity while preserving privacy, prior research has suggested calculating distances between model weights \cite{Kairouz2021}. 
However, as model complexity grows, direct weight analysis becomes impractically burdensome. 
Drawing from clustering methods in \cite{Ouyang2021}, we instead compare client model predictions using Kullback-Leibler (KL) Divergence $\text{D}_{\text{KL}}(\cdot)$, defined as:
\begin{equation}
\text{D}_{\text{KL}}(\mathbf{P} || \mathbf{Q}) = \frac{1}{N}\sum\nolimits_{n = 0}^{N-1}P_{n}\log \left(\frac{P_{n}}{Q_{n}}\right)
\end{equation}
where $P_{n}$ ($Q_{n}$) represents the $n$-th element of the probability distribution $\mathbf{P}$ ($\mathbf{Q}$).
KL Divergence is non-negative, with higher values indicating greater divergence.

To quantitatively evaluate the similarity in data distribution between client $k$ and client $j$, we first calculate the model distances by the following equation:
\begin{equation} 
d_{k,j} = \mathbb{E}_{x_i \sim \mathbb{D}_{s, u}} [\text{D}_{\text{KL}}\left(\sigma(f(\theta_{k}^{*}, x_i)) \: || \: \sigma(f(\theta_{j}^{*}, x_i))\right)]
\end{equation}

Taking into account the characteristics of KL Divergence, we apply an exponential decay function to the calculated distances, which effectively decreases the impact score as the divergence increases, then normalize the model similarity, formulated as:
\begin{equation}
\label{eq:s}
c_{k, j} = \frac{\exp(-d_{k,j})}{\sum_{j}\exp(-d_{k,j})}
\end{equation}

We utilize this normalized similarity score as the compatibility score between clients, which allows for a nuanced analysis of client heterogeneity in FL environments. 

\noindent\textbf{Value:}
This metric measures the importance of each client model in enhancing the server model’s performance.
We recognize that FL is characterized by allowing clients to access the complete server model while keeping local datasets private.
Research by Goetz et al. \cite{Goetz2019} and Cho et al. \cite{Cho2022}, using experimental and theoretical analysis respectively, demonstrated that lower server model accuracy on a local dataset indicates significant potential for the corresponding local model to enhance the server model's performance.
Therefore, our value function prioritizes clients whose data is expected to provide the greatest performance boost in each training iteration, ensuring efficient and impactful model development.

For client $k$, the value is quantified as follows \cite{Cho2022}: 
\begin{equation}
\label{eq:v}
v_{k}^{t} = \mathcal{L}(\mathbb{D}_{k}, \theta_s^{t-1})
\end{equation}

\noindent\textbf{Attenton score:}
The score for client model $\theta_{k}$ is determined using compatibility scores and values:
\begin{equation}
\mathcal{S}(\theta_{k}^{t}) = \sum\nolimits_{j}c_{k, j}v_{j}^{t}
\end{equation}
Here, $c_{k, j}$, generated by Eq.~(\ref{eq:s}), represents the compatibility score between client $k$ and client $j$, and $v_j^t$ from Eq.~(\ref{eq:v}), denotes client $j$'s contributions in enhancing the server model.

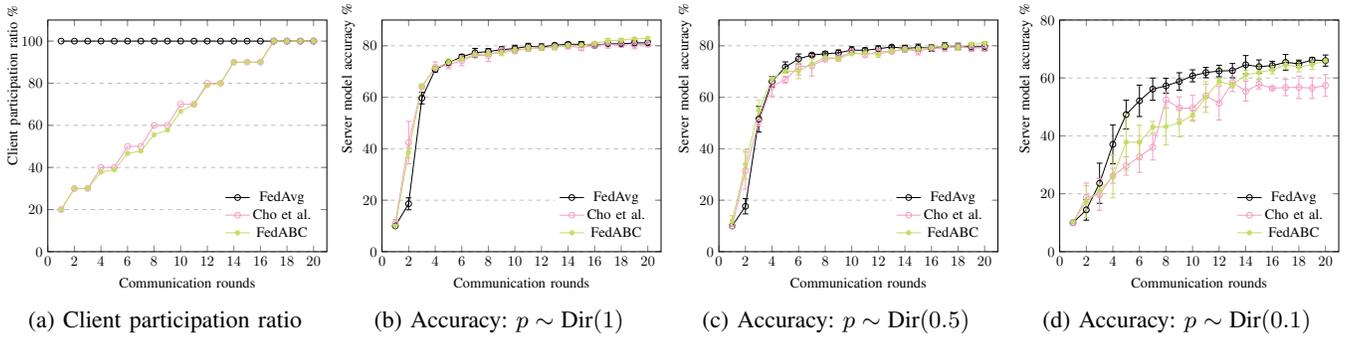
\begin{figure*}[t]
\centering
\begin{subfigure}{.245\linewidth}
\resizebox{\textwidth}{!}{
\begin{tikzpicture}
\begin{axis}[
    xlabel={Communication rounds},
    ylabel={Client participation ratio \%},
    ylabel style={at={(axis description cs:0.1,0.72)}, anchor=south}, 
    xmin=0, xmax=21,
    ymin=0, ymax=110,
    xtick={0,2,4,6,8,10,12,14,16,18,20},
    ytick={0,20,40,60,80, 100},
    legend style={draw=none},
    legend pos=south east,
    legend columns=1, 
    ymajorgrids=true,
    grid style=dashed,
]

\addplot [
    color=black,
    mark=o,
    error bars/.cd,
    y dir=both,
    y explicit
    ]
    coordinates{
    (1, 100)(2, 100)(3, 100)(4, 100) (5, 100)(6, 100)(7, 100)(8, 100)(9, 100)(10, 100)(11, 100)(12, 100)(13, 100)(14, 100)(15, 100)(16, 100)(17, 100)(18, 100)(19, 100)(20, 100)
    }; 
\addlegendentry{FedAvg}

\addplot[
    color=Lavender,
    mark=o,
    ]
    coordinates {
    (1, 20)(2, 30)(3, 30)(4, 40) (5, 40)(6, 50)(7, 50)(8, 60)(9, 60)(10, 70)(11, 70)(12, 80)(13, 80)(14, 90)(15, 90)(16, 90)(17, 100)(18, 100)(19, 100)(20, 100)
    };
\addlegendentry{Cho et al.} 
         
\addplot[
    color=SpringGreen,
    mark=10-pointed star,
    ]
    coordinates {
    (1,20)(2,30)(3,30)(4,37.77777778)(5,38.88888889)(6,46.66666667)(7,47.77777778)(8,55.55555556)(9,57.77777778)(10,66.66666667)(11,70)(12,78.88888889)(13,80)(14,90)(15,90)(16,90)(17,100)(18,100)(19,100)(20,100)
    };
\addlegendentry{FedABC} 

\end{axis}
\end{tikzpicture}
}
\caption{Client participation ratio}
\label{fig:com_part}
\end{subfigure}%
\begin{subfigure}{.245\linewidth}
\resizebox{\textwidth}{!}{
\begin{tikzpicture}
\begin{axis}[
    xlabel={Communication rounds},
    ylabel={Server model accuracy \%},
    ylabel style={at={(axis description cs:0.1,0.75)}, anchor=south}, 
    xmin=0, xmax=21,
    xtick={0,2,4,6,8,10,12,14,16,18,20},
    ymin=0, ymax=90,
    ytick={0,20,40, 60, 80},
    error bars/.cd,
    legend style={draw=none},
    legend pos=south east,
    legend columns=1, 
    ymajorgrids=true,
    grid style=dashed,
]

\addplot+[
     color=black,
     mark=o,
        error bars/.cd,
        y dir=both,
        y explicit
    ] coordinates {
        (1, 10.0)        +- (0, 0.0)
        (2, 18.6)        +- (0, 2.31516738)
        (3, 59.66666667) +- (0, 2.2395436)
        (4, 70.76666667) +- (0, 1.02089286)
        (5, 73.4)        +- (0, 0.45460606)
        (6, 75.73333333) +- (0, 0.87305339)
        (7, 77.3)        +- (0, 1.68720676)
        (8, 77.76666667) +- (0, 1.15566239)
        (9, 78.5)        +- (0, 1.1045361)
        (10, 78.93333333) +- (0, 1.26578917)
        (11, 79.76666667) +- (0, 1.18977122)
        (12, 79.53333333) +- (0, 1.22292909)
        (13, 80.23333333) +- (0, 0.55577773)
        (14, 80.6)        +- (0, 0.69761498)
        (15, 80.56666667) +- (0, 1.02089286)
        (16, 80.23333333) +- (0, 0.69442222)
        (17, 80.76666667) +- (0, 0.88065632)
        (18, 80.76666667) +- (0, 0.51854497)
        (19, 81.1)        +- (0, 1.0033278)
        (20, 81.16666667) +- (0, 0.81785628)
    };
    \addlegendentry{FedAvg}

 \addplot+[
        color=Lavender,
        mark=o,
        error bars/.cd,
        y dir=both,
        y explicit
    ] coordinates {
        (1, 11.03333333) +- (0, 1.46135401)
        (2, 42.4)        +- (0, 8.2659946)
        (3, 64.1)        +- (0, 0.37416574)
        (4, 71.63333333) +- (0, 2.10449255)
        (5, 72.6)        +- (0, 1.51217283)
        (6, 74.2)        +- (0, 1.91311265)
        (7, 76.26666667) +- (0, 0.32998316)
        (8, 76.26666667) +- (0, 2.31708629)
        (9, 78.26666667) +- (0, 0.80138769)
        (10, 78.0)       +- (0, 0.53541261)
        (11, 78.96666667) +- (0, 0.77172246)
        (12, 79.26666667) +- (0, 1.02089286)
        (13, 79.6)       +- (0, 0.98994949)
        (14, 79.9)       +- (0, 1.1343133)
        (15, 79.3)       +- (0, 1.20277457)
        (16, 80.26666667) +- (0, 1.18977122)
        (17, 80.46666667) +- (0, 1.40791414)
        (18, 80.4)       +- (0, 0.92736185)
        (19, 80.33333333) +- (0, 1.48174072)
        (20, 80.56666667) +- (0, 0.8993825)
    };
    \addlegendentry{Cho et al.}

\addplot+[
     color=SpringGreen,
     mark=10-pointed star,
    error bars/.cd,
    y dir=both,
    y explicit
    ] coordinates {
        (1, 10.0)        +- (0, 0.0)
        (2, 38.5)        +- (0, 1.92006944)
        (3, 64.33333333) +- (0, 0.83798701)
        (4, 71.33333333) +- (0, 0.16996732)
        (5, 73.6)        +- (0, 1.34907376)
        (6, 74.83333333) +- (0, 0.65489609)
        (7, 76.56666667) +- (0, 0.54365021)
        (8, 76.83333333) +- (0, 1.02740233)
        (9, 77.13333333) +- (0, 0.98432154)
        (10, 78.2)       +- (0, 1.22474487)
        (11, 79.03333333) +- (0, 1.22836838)
        (12, 79.13333333) +- (0, 0.96724121)
        (13, 79.4)       +- (0, 1.0033278)
        (14, 80.13333333) +- (0, 0.73181661)
        (15, 80.16666667) +- (0, 0.82192187)
        (16, 81.0)       +- (0, 0.50990195)
        (17, 81.93333333) +- (0, 0.92855922)
        (18, 82.23333333) +- (0, 0.61824123)
        (19, 82.7)       +- (0, 0.42426407)
        (20, 82.86666667) +- (0, 0.54365021)
    };
    \addlegendentry{FedABC}
    
\end{axis}
\end{tikzpicture}
}
\caption{Accuracy: $p \sim \text{Dir}(1)$}
\label{fig:com_1}
\end{subfigure}%
\hspace{0.1em}
\begin{subfigure}{.245\linewidth}
\resizebox{\textwidth}{!}{
\begin{tikzpicture}
\begin{axis}[
    xlabel={Communication rounds},
    ylabel={Server model accuracy \%},
    ylabel style={at={(axis description cs:0.1,0.75)}, anchor=south}, 
    xmin=0, xmax=21,
    xtick={ 0,2,4,6,8,10,12,14,16,18,20},
    ymin=0, ymax=90,
    ytick={0,20,40, 60, 80},
    error bars/.cd,
    legend style={draw=none},
    legend pos=south east,
    legend columns=1, 
    ymajorgrids=true,
    grid style=dashed,
]

\addplot+[
     color=black,
     mark=o,
        error bars/.cd,
        y dir=both,
        y explicit
    ] coordinates {
        (1, 10.0)        +- (0, 0.0)
        (2, 17.63333333) +- (0, 2.94882274)
        (3, 51.5)        +- (0, 4.98062914)
        (4, 66.06666667) +- (0, 2.02703944)
        (5, 71.73333333) +- (0, 2.01549553)
        (6, 74.9)        +- (0, 1.92527054)
        (7, 76.43333333) +- (0, 0.601849)
        (8, 76.9)        +- (0, 0.7788881)
        (9, 77.26666667) +- (0, 1.11455023)
        (10, 78.33333333) +- (0, 1.302135)
        (11, 78.23333333) +- (0, 0.94633797)
        (12, 78.86666667) +- (0, 1.11455023)
        (13, 79.53333333) +- (0, 0.65996633)
        (14, 79.0)        +- (0, 1.21928941)
        (15, 79.3)        +- (0, 1.35892114)
        (16, 79.1)        +- (0, 1.25698051)
        (17, 79.86666667) +- (0, 1.34742553)
        (18, 79.73333333) +- (0, 1.27627931)
        (19, 79.53333333) +- (0, 1.44299072)
        (20, 79.7)        +- (0, 1.42361043)
    };
    \addlegendentry{FedAvg}

     \addplot+[
        color=Lavender,
        mark=o,
        error bars/.cd,
        y dir=both,
        y explicit
    ] coordinates {
        (1, 10.0)         +- (0, 0.0)
        (2, 31.53333333)  +- (0, 7.22511053)
        (3, 50.5)         +- (0, 1.56418243)
        (4, 64.23333333)  +- (0, 3.7043518)
        (5, 66.83333333)  +- (0, 1.32748718)
        (6, 71.83333333)  +- (0, 3.20659044)
        (7, 72.33333333)  +- (0, 4.06147209)
        (8, 74.73333333)  +- (0, 0.65996633)
        (9, 75.33333333)  +- (0, 1.11155547)
        (10, 77.6)         +- (0, 0.37416574)
        (11, 76.43333333)  +- (0, 0.82192187)
        (12, 77.7)         +- (0, 1.31402689)
        (13, 77.76666667)  +- (0, 0.55577773)
        (14, 78.6)         +- (0, 1.0984838)
        (15, 77.93333333)  +- (0, 1.48174072)
        (16, 78.96666667)  +- (0, 0.57348835)
        (17, 79.1)         +- (0, 1.17756812)
        (18, 79.66666667)  +- (0, 1.23917535)
        (19, 79.13333333)  +- (0, 1.24988888)
        (20, 79.2)         +- (0, 1.34907376)
    };
    \addlegendentry{Cho et al.}

 \addplot+[
     color=SpringGreen,
     mark=10-pointed star,
    error bars/.cd,
    y dir=both,
    y explicit
    ] coordinates {
        (1, 12.0)  +- (0, 2.0)
        (2, 33.9)  +- (0, 5.7)
        (3, 54.9)  +- (0, 4.1)
        (4, 67.1)  +- (0, 0.1)
        (5, 70.05) +- (0, 0.75)
        (6, 70.15) +- (0, 2.95)
        (7, 73.0)  +- (0, 1.9)
        (8, 75.55) +- (0, 0.05)
        (9, 75.05) +- (0, 1.05)
        (10, 76.85) +- (0, 0.05)
        (11, 77.3) +- (0, 0.3)
        (12, 76.65) +- (0, 1.15)
        (13, 78.0)  +- (0, 0.8)
        (14, 78.55) +- (0, 0.65)
        (15, 78.75) +- (0, 0.05)
        (16, 79.2)  +- (0, 0.3)
        (17, 79.95) +- (0, 1.05)
        (18, 79.35) +- (0, 1.15)
        (19, 80.6)  +- (0, 0.7)
        (20, 80.93)  +- (0, 0.7)
    };
    \addlegendentry{FedABC}

\end{axis}
\end{tikzpicture}
}
\caption{Accuracy: $p \sim \text{Dir}(0.5)$}
\label{fig:com_0.5}
\end{subfigure}%
\hspace{0.25em}
\begin{subfigure}{.245\linewidth}
\resizebox{\textwidth}{!}{
\begin{tikzpicture}
\begin{axis}[
    xlabel={Communication rounds},
    ylabel={Server model accuracy \%},
    ylabel style={at={(axis description cs:0.1,0.75)}, anchor=south}, 
    xmin=0, xmax=21,
    ymin=0, ymax=80,
    xtick={ 0,2,4,6,8,10,12,14,16,18,20},
    ytick={0,20,40,60,80},
    legend style={draw=none},
    legend pos=south east,
    legend columns=1, 
    ymajorgrids=true,
    grid style=dashed,
]

\addplot+[
     color=black,
     mark=o,
        error bars/.cd,
        y dir=both,
        y explicit
    ] coordinates {
        (1, 10.0)        +- (0, 0.0)
        (2, 14.43333333) +- (0, 3.5537148)
        (3, 23.63333333) +- (0, 6.93653772)
        (4, 37.06666667) +- (0, 6.71532245)
        (5, 47.36666667) +- (0, 4.97013302)
        (6, 52.1)        +- (0, 5.41664103)
        (7, 56.13333333) +- (0, 3.84216374)
        (8, 57.26666667) +- (0, 2.59786237)
        (9, 58.8)        +- (0, 3.00111091)
        (10, 60.73333333) +- (0, 2.07417989)
        (11, 61.86666667) +- (0, 1.79133718)
        (12, 62.4)        +- (0, 1.98662192)
        (13, 62.53333333) +- (0, 2.46080384)
        (14, 64.56666667) +- (0, 3.1678945)
        (15, 63.9)        +- (0, 2.33380948)
        (16, 64.26666667) +- (0, 1.48174072)
        (17, 65.4)        +- (0, 2.69938265)
        (18, 64.9)        +- (0, 1.26754356)
        (19, 66.2)        +- (0, 0.89814624)
        (20, 66.0)        +- (0, 1.92006944)
    };
    \addlegendentry{FedAvg}

    \addplot+[
        color=Lavender,
        mark=o,
        error bars/.cd,
        y dir=both,
        y explicit
    ] coordinates {
        (1, 10.0)        +- (0, 0.0)
        (2, 18.86666667) +- (0, 4.86506823)
        (3, 19.76666667) +- (0, 5.47803696)
        (4, 26.2)        +- (0, 2.596151)
        (5, 29.6)        +- (0, 3.18433667)
        (6, 32.8)        +- (0, 5.41356321)
        (7, 36.06666667) +- (0, 4.40782133)
        (8, 52.43333333) +- (0, 2.71825107)
        (9, 49.6)        +- (0, 3.906405)
        (10, 49.56666667) +- (0, 4.54043561)
        (11, 53.86666667) +- (0, 3.94658784)
        (12, 51.26666667) +- (0, 5.77369514)
        (13, 58.46666667) +- (0, 3.20659044)
        (14, 55.43333333) +- (0, 3.39542175)
        (15, 57.83333333) +- (0, 1.62138487)
        (16, 56.43333333) +- (0, 0.57927157)
        (17, 56.7)        +- (0, 2.81780056)
        (18, 56.76666667) +- (0, 3.89643712)
        (19, 56.5)        +- (0, 3.60647566)
        (20, 57.4)        +- (0, 3.76651917)
    };
    \addlegendentry{Cho et al.}

    \addplot+[
     color=SpringGreen,
     mark=10-pointed star,
    error bars/.cd,
    y dir=both,
    y explicit
    ] coordinates {
        (1, 10.2)  +- (0, 0.2)
        (2, 17.3)  +- (0, 5.5)
        (3, 21.0)  +- (0, 2.1)
        (4, 26.05) +- (0, 7.45)
        (5, 37.85) +- (0, 9.75)
        (6, 37.75) +- (0, 5.95)
        (7, 43.05) +- (0, 1.95)
        (8, 43.2)  +- (0, 6.3)
        (9, 44.5)  +- (0, 4.9)
        (10, 47.15) +- (0, 1.65)
        (11, 53.4)  +- (0, 5.3)
        (12, 58.8)  +- (0, 1.1)
        (13, 57.55) +- (0, 0.95)
        (14, 61.05) +- (0, 1.65)
        (15, 61.95) +- (0, 2.55)
        (16, 62.75) +- (0, 1.35)
        (17, 64.5)  +- (0, 1.4)
        (18, 63.8)  +- (0, 1.7)
        (19, 64.6)  +- (0, 1.6)
        (20, 65.95) +- (0, 0.45)  
    };
    \addlegendentry{FedABC}
    
\end{axis}
\end{tikzpicture}
}
\caption{Accuracy: $p \sim \text{Dir}(0.1)$}
\label{fig:com_0.1}
\end{subfigure}%
\caption {
Performance comparison with baselines under various data distribution heterogeneity. 
}
\label{fig:com}
\end{figure*}
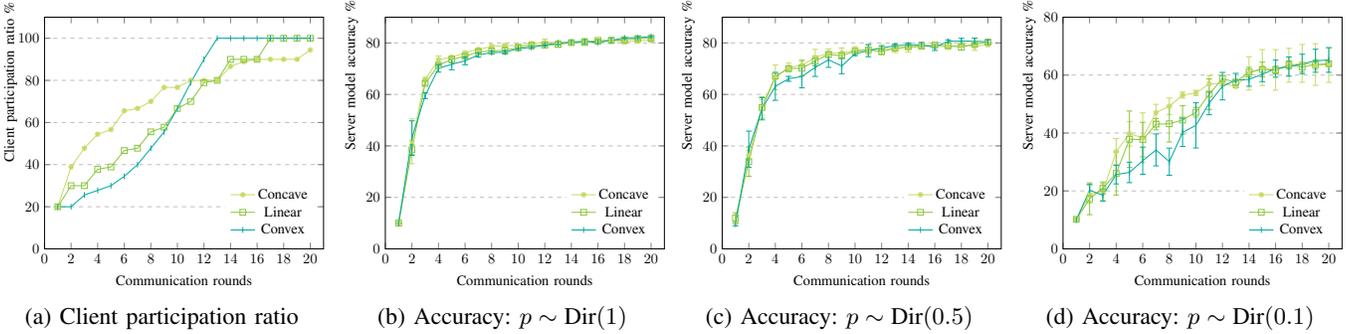
\begin{figure*}[t]
\centering
\begin{subfigure}{.245\linewidth}
\resizebox{\textwidth}{!}{
\begin{tikzpicture}
\begin{axis}[
    xlabel={Communication rounds},
    ylabel={Client participation ratio \%},
    ylabel style={at={(axis description cs:0.1,0.72)}, anchor=south}, 
    xmin=0, xmax=21,
    ymin=0, ymax=110,
    xtick={0,2,4,6,8,10,12,14,16,18,20},
    ytick={0,20,40,60,80, 100},
    legend style={draw=none},
    legend pos=south east,
    legend columns=1, 
    ymajorgrids=true,
    grid style=dashed,
]

\addplot[
    color=SpringGreen,
    mark=10-pointed star,
    ]
    coordinates {
    (1, 20)(2, 38.88888889)(3, 47.77777778)(4, 54.44444444) (5, 56.66666667)(6, 65.55555556)(7, 66.66666667)(8, 70)(9, 76.66666667)(10, 76.66666667)(11, 80)(12, 80)(13, 80)(14, 86.66666667)(15, 88.88888889)(16, 90)(17, 90)(18, 90)(19, 90)(20, 94.44444444)
    };
    \addlegendentry{Concave}

\addplot[
    color=LimeGreen,
    mark=square,
    ]
    coordinates {
    (1,20)(2,30)(3,30)(4,37.77777778)(5,38.88888889)(6,46.66666667)(7,47.77777778)(8,55.55555556)(9,57.77777778)(10,66.66666667)(11,70)(12,78.88888889)(13,80)(14,90)(15,90)(16,90)(17,100)(18,100)(19,100)(20,100)
    };
    \addlegendentry{Linear}

\addplot[
    color=Emerald,
    mark=|,
    ]
    coordinates {
    (1, 20) (2, 20) (3, 25.55555556) (4, 27.77777778) (5, 30) (6, 34.44444444)
    (7, 40) (8, 47.77777778) (9, 55.55555556) (10, 66.66666667) (11, 78.88888889) (12, 90)
    (13, 100) (14, 100) (15, 100) (16, 100) (17, 100)
    (18, 100) (19, 100) (20, 100)
    };
    \addlegendentry{Convex}

\end{axis}
\end{tikzpicture}
}
\caption{Client participation ratio}
\label{fig:part}
\end{subfigure}%
\hspace{0.25em}
\begin{subfigure}{.245\linewidth}
\resizebox{\textwidth}{!}{
\begin{tikzpicture}
\begin{axis}[
    xlabel={Communication rounds},
    ylabel={Server model accuracy \%},
    ylabel style={at={(axis description cs:0.1,0.75)}, anchor=south}, 
    xmin=0, xmax=21,
    xtick={ 0,2,4,6,8,10,12,14,16,18,20},
    ymin=0, ymax=90,
    ytick={0,20,40, 60, 80},
    error bars/.cd,
   legend style={draw=none},
    legend pos=south east,
    legend columns=1, 
    ymajorgrids=true,
    grid style=dashed,
]
    
 \addplot+[
     color=SpringGreen,
     mark=10-pointed star,
    error bars/.cd,
    y dir=both,
    y explicit
    ] coordinates {
        (1, 10.0)        +- (0, 0.0)
        (2, 41.8)        +- (0, 8.70210702)
        (3, 65.86666667) +- (0, 0.69442222)
        (4, 73.5)        +- (0, 1.45143607)
        (5, 74.53333333) +- (0, 0.26246693)
        (6, 76.13333333) +- (0, 0.49216077)
        (7, 77.56666667) +- (0, 0.3681787)
        (8, 78.56666667) +- (0, 0.75865378)
        (9, 79.1)        +- (0, 0.50990195)
        (10, 78.86666667) +- (0, 0.54365021)
        (11, 79.7)       +- (0, 0.69761498)
        (12, 80.3)       +- (0, 1.26754356)
        (13, 80.26666667) +- (0, 0.26246693)
        (14, 80.03333333) +- (0, 0.74087036)
        (15, 80.7)       +- (0, 1.08012345)
        (16, 80.3)       +- (0, 0.7788881)
        (17, 81.03333333) +- (0, 0.95335664)
        (18, 80.43333333) +- (0, 0.84983659)
        (19, 80.83333333) +- (0, 0.90308115)
        (20, 81.03333333) +- (0, 0.71336449)
    };
    \addlegendentry{Concave}

 \addplot+[
        color=LimeGreen,
        mark=square,
        error bars/.cd,
        y dir=both,
        y explicit
    ] coordinates {
        (1, 10.0)        +- (0, 0.0)
        (2, 38.5)        +- (0, 1.92006944)
        (3, 64.33333333) +- (0, 0.83798701)
        (4, 71.33333333) +- (0, 0.16996732)
        (5, 73.6)        +- (0, 1.34907376)
        (6, 74.83333333) +- (0, 0.65489609)
        (7, 76.56666667) +- (0, 0.54365021)
        (8, 76.83333333) +- (0, 1.02740233)
        (9, 77.13333333) +- (0, 0.98432154)
        (10, 78.2)       +- (0, 1.22474487)
        (11, 79.03333333) +- (0, 1.22836838)
        (12, 79.13333333) +- (0, 0.96724121)
        (13, 79.4)       +- (0, 1.0033278)
        (14, 80.13333333) +- (0, 0.73181661)
        (15, 80.16666667) +- (0, 0.82192187)
        (16, 81.0)       +- (0, 0.50990195)
        (17, 80.93333333) +- (0, 0.92855922)
        (18, 81.23333333) +- (0, 0.61824123)
        (19, 81.7)       +- (0, 0.42426407)
        (20, 81.86666667) +- (0, 0.54365021)
    };
    \addlegendentry{Linear}

 \addplot+[
        color=Emerald,
        mark=|,
        error bars/.cd,
        y dir=both,
        y explicit
    ] coordinates {
        (1, 10.0)        +- (0, 0.0)
        (2, 43.03333333) +- (0, 6.71184691)
        (3, 59.53333333) +- (0, 1.14406682)
        (4, 70.16666667) +- (0, 1.50628314)
        (5, 71.7)        +- (0, 2.11817532)
        (6, 73.06666667) +- (0, 1.53260852)
        (7, 75.3)        +- (0, 0.73484692)
        (8, 76.26666667) +- (0, 0.62360956)
        (9, 76.36666667) +- (0, 0.65996633)
        (10, 77.83333333) +- (0, 0.28674418)
        (11, 78.26666667) +- (0, 0.38586123)
        (12, 79.2)       +- (0, 0.35590261)
        (13, 79.9)       +- (0, 0.65319726)
        (14, 80.36666667) +- (0, 0.47140452)
        (15, 80.63333333) +- (0, 0.82596745)
        (16, 80.2)       +- (0, 0.58878406)
        (17, 81.0)       +- (0, 1.04243305)
        (18, 82.0)       +- (0, 0.29439203)
        (19, 82.06666667) +- (0, 0.54365021)
        (20, 82.46666667) +- (0, 0.37712362)
    };
    \addlegendentry{Convex}
    
\end{axis}
\end{tikzpicture}
}
\caption{Accuracy: $p \sim \text{Dir}(1)$}
\label{fig:1}
\end{subfigure}%
\hspace{0.1em}
\begin{subfigure}{.245\linewidth}
\resizebox{\textwidth}{!}{
\begin{tikzpicture}
\begin{axis}[
    xlabel={Communication rounds},
    ylabel={Server model accuracy \%},
    ylabel style={at={(axis description cs:0.1,0.75)}, anchor=south}, 
    xmin=0, xmax=21,
    xtick={ 0,2,4,6,8,10,12,14,16,18,20},
    ymin=0, ymax=90,
    ytick={0,20,40, 60, 80},
    error bars/.cd,
    legend style={draw=none},
    legend pos=south east,
    legend columns=1, 
    ymajorgrids=true,
    grid style=dashed,
]

 \addplot+[
     color=SpringGreen,
     mark=10-pointed star,
    error bars/.cd,
    y dir=both,
    y explicit
    ] coordinates {
        (1, 10.0)        +- (0, 0.0)
        (2, 36.4)        +- (0, 1.80554701)
        (3, 54.96666667) +- (0, 3.02691629)
        (4, 66.6)        +- (0, 4.82493523)
        (5, 70.43333333) +- (0, 1.73076733)
        (6, 71.4)        +- (0, 2.16487105)
        (7, 74.33333333) +- (0, 3.1678945)
        (8, 76.03333333) +- (0, 1.68193011)
        (9, 76.2)        +- (0, 1.84028983)
        (10, 77.03333333) +- (0, 1.55205956)
        (11, 77.46666667) +- (0, 1.93964487)
        (12, 77.9)       +- (0, 1.47196014)
        (13, 77.6)       +- (0, 1.34907376)
        (14, 78.03333333) +- (0, 1.56914697)
        (15, 78.76666667) +- (0, 1.15566239)
        (16, 78.86666667) +- (0, 0.95335664)
        (17, 78.4)       +- (0, 0.81649658)
        (18, 79.4)       +- (0, 0.66833126)
        (19, 78.6)       +- (0, 1.49666295)
        (20, 79.43333333) +- (0, 0.84983659)
    };
    \addlegendentry{Concave}

 \addplot+[
        color=LimeGreen,
        mark=square,
        error bars/.cd,
        y dir=both,
        y explicit
    ] coordinates {
        (1, 12.0)  +- (0, 2.0)
        (2, 33.9)  +- (0, 5.7)
        (3, 54.9)  +- (0, 4.1)
        (4, 67.1)  +- (0, 0.1)
        (5, 70.05) +- (0, 0.75)
        (6, 70.15) +- (0, 2.95)
        (7, 73.0)  +- (0, 1.9)
        (8, 75.55) +- (0, 0.05)
        (9, 75.05) +- (0, 1.05)
        (10, 76.85) +- (0, 0.05)
        (11, 77.3) +- (0, 0.3)
        (12, 76.65) +- (0, 1.15)
        (13, 78.0)  +- (0, 0.8)
        (14, 78.55) +- (0, 0.65)
        (15, 78.75) +- (0, 0.05)
        (16, 79.2)  +- (0, 0.3)
        (17, 78.95) +- (0, 1.05)
        (18, 78.35) +- (0, 1.15)
        (19, 79.6)  +- (0, 0.7)
        (20, 80.3)  +- (0, 0.7)
    };
    \addlegendentry{Linear}

 \addplot+[
        color=Emerald,
        mark=|,
        error bars/.cd,
        y dir=both,
        y explicit
    ] coordinates {
        (1, 9.56666667)   +- (0, 0.61282588)
        (2, 38.73333333)  +- (0, 7.00682207)
        (3, 54.43333333)  +- (0, 4.32229363)
        (4, 63.13333333)  +- (0, 5.43098109)
        (5, 66.06666667)  +- (0, 0.94633797)
        (6, 66.96666667)  +- (0, 4.41008944)
        (7, 70.5)         +- (0, 3.50523418)
        (8, 73.46666667)  +- (0, 2.96348144)
        (9, 71.03333333)  +- (0, 2.97806798)
        (10, 75.83333333) +- (0, 0.75865378)
        (11, 77.0)        +- (0, 2.34236348)
        (12, 78.06666667) +- (0, 0.74087036)
        (13, 78.9)        +- (0, 0.65319726)
        (14, 79.26666667) +- (0, 0.71336449)
        (15, 79.23333333) +- (0, 0.96724121)
        (16, 78.3)        +- (0, 1.21928941)
        (17, 80.66666667) +- (0, 0.52493386)
        (18, 80.7)        +- (0, 1.0984838)
        (19, 80.73333333) +- (0, 1.13235252)
        (20, 80.4)        +- (0, 0.84852814)
    };
    \addlegendentry{Convex}

\end{axis}
\end{tikzpicture}
}
\caption{Accuracy: $p \sim \text{Dir}(0.5)$}
\label{fig:0.5}
\end{subfigure}%
\hspace{0.25em}
\begin{subfigure}{.245\linewidth}
\resizebox{\textwidth}{!}{
\begin{tikzpicture}
\begin{axis}[
    xlabel={Communication rounds},
    ylabel={Server model accuracy \%},
    ylabel style={at={(axis description cs:0.1,0.75)}, anchor=south}, 
    xmin=0, xmax=21,
    ymin=0, ymax=80,
    xtick={ 0,2,4,6,8,10,12,14,16,18,20},
    ytick={0,20,40,60,80},
    legend style={draw=none},
    legend pos=south east,
    legend columns=1, 
    ymajorgrids=true,
    grid style=dashed,
]

 \addplot+[
     color=SpringGreen,
     mark=10-pointed star,
    error bars/.cd,
    y dir=both,
    y explicit
    ] coordinates {
        (1, 10.2)        +- (0, 0.28284271)
        (2, 18.8)        +- (0, 3.02324329)
        (3, 19.76666667) +- (0, 2.91471363)
        (4, 33.56666667) +- (0, 4.53455866)
        (5, 39.73333333) +- (0, 4.24918293)
        (6, 38.4)        +- (0, 8.56076321)
        (7, 47.03333333) +- (0, 2.868604)
        (8, 49.23333333) +- (0, 2.85228953)
        (9, 53.1)        +- (0, 1.06770783)
        (10, 53.83333333) +- (0, 0.92855922)
        (11, 56.96666667) +- (0, 2.89405521)
        (12, 57.2)       +- (0, 0.92014492)
        (13, 56.06666667) +- (0, 0.3681787)
        (14, 60.6)        +- (0, 5.75557701)
        (15, 62.6)        +- (0, 6.21771662)
        (16, 61.8)        +- (0, 7.00142843)
        (17, 63.4)        +- (0, 5.84465568)
        (18, 64.06666667) +- (0, 6.53826345)
        (19, 63.66666667) +- (0, 7.1908893)
        (20, 63.43333333) +- (0, 5.92864984) 
    };
    \addlegendentry{Concave}

 \addplot+[
        color=LimeGreen,
        mark=square,
        error bars/.cd,
        y dir=both,
        y explicit
    ] coordinates {
        (1, 10.2)  +- (0, 0.2)
        (2, 17.3)  +- (0, 5.5)
        (3, 21.0)  +- (0, 2.1)
        (4, 26.05) +- (0, 7.45)
        (5, 37.85) +- (0, 9.75)
        (6, 37.75) +- (0, 5.95)
        (7, 43.05) +- (0, 1.95)
        (8, 43.2)  +- (0, 6.3)
        (9, 44.5)  +- (0, 4.9)
        (10, 47.15) +- (0, 1.65)
        (11, 53.4)  +- (0, 5.3)
        (12, 58.8)  +- (0, 1.1)
        (13, 57.55) +- (0, 0.95)
        (14, 61.05) +- (0, 1.65)
        (15, 61.95) +- (0, 2.55)
        (16, 61.75) +- (0, 1.35)
        (17, 63.5)  +- (0, 1.4)
        (18, 62.8)  +- (0, 1.7)
        (19, 63.6)  +- (0, 1.6)
        (20, 63.95) +- (0, 0.45)
    };
    \addlegendentry{Linear}

 \addplot+[
        color=Emerald,
        mark=|,
        error bars/.cd,
        y dir=both,
        y explicit
    ] coordinates {
        (1, 10.0)        +- (0, 0.0)
        (2, 20.26666667) +- (0, 1.98718114)
        (3, 18.5)        +- (0, 2.08326667)
        (4, 25.66666667) +- (0, 3.23144275)
        (5, 26.4)        +- (0, 3.53930313)
        (6, 30.43333333) +- (0, 4.69491451)
        (7, 34.16666667) +- (0, 5.51986312)
        (8, 30.1)        +- (0, 4.68472696)
        (9, 40.23333333) +- (0, 4.87601841)
        (10, 42.63333333) +- (0, 7.79843289)
        (11, 50.3)       +- (0, 3.98580816)
        (12, 56.2)       +- (0, 4.74201083)
        (13, 58.2)       +- (0, 2.40554914)
        (14, 58.53333333) +- (0, 2.37954244)
        (15, 60.3)       +- (0, 2.54689353)
        (16, 62.23333333) +- (0, 2.95334086)
        (17, 62.93333333) +- (0, 3.3159547)
        (18, 63.86666667) +- (0, 3.35194802)
        (19, 64.96666667) +- (0, 3.96680672)
        (20, 65.23333333) +- (0, 4.26249796)
    };
    \addlegendentry{Convex}
    
\end{axis}
\end{tikzpicture}
}
\caption{Accuracy: $p \sim \text{Dir}(0.1)$}
\label{fig:0.1}
\end{subfigure}%
\caption {
Performance analysis under different threshold designs. 
}
\label{fig:cifar10}
\end{figure*}

\subsection{Client Selection and Model Aggregation}

For global round $t$, Eq.~(\ref{eq:optt}) is rewritten as:
\begin{equation}
\label{eq:final}
       \max  \sum_{k=0}^{K-1} m_{k}^{t} (\eta^t \mathcal{S}(\theta_{k}^{t}) - \lambda)
\end{equation}

The client selection decision can be easily obtained by setting $m_{k}^{t} = 1$ when the corresponding $ \mathcal{S}(\theta_{k}^{t}) > \lambda / \eta^t$, $m_{k}^{t} = 0$ otherwise.
This approach prioritizes clients who are most likely to contribute to the global model significantly, optimizing the training process in each global round.
The whole selection procedure is denoted as $\mathcal{G}(\cdot)$ in Fig. ~\ref{fig:arch}.

We adopt the model aggregation method described in Eq.~(\ref{eq:a}), where the weight of each client $k$, denoted as $w_k$ is defined by normalized score values, defined as:
\begin{equation}
\label{eq:weight}
      w_k = \frac{m_{k}^{t} \cdot \mathcal{S}(\theta_{k})}{\sum_{k=0}^{K-1} m_{k}^{t} \cdot \mathcal{S}(\theta_{k})}
\end{equation}
This normalization ensures that the scores are proportionately scaled to reflect the relative importance of each client.

\section{NUMERICAL EVALUATION}

\subsection{Experiment Setup}
\noindent\textbf{Dataset settings:}
We conduct image classification tasks using the CIFAR-10 dataset, which contains 60,000 color images across 10 classes. 
We randomly divide each dataset into two parts: a server dataset and a client dataset. 
The public dataset, consisting of $5,000$ samples, has all labels removed and obtained by the server.
To simulate different distribution heterogeneity in the client dataset, we use Dirichlet distribution: 
$p \sim \text{Dir}(\alpha)$, where for each class, the proportions $p$ of data samples for each client are sampled from a Dirichlet distribution, with a lower $\alpha$ value indicating greater heterogeneity.

%

\noindent\textbf{Model and parameters settings:}
The system includes $K = 10$ clients and one server, all using the ResNet20 model. 
We set the global round $T$ as 20, and for each model training, we run $20$ local epochs. 
The batch size is set to $64$ and the learning rate is $0.001$.

\noindent\textbf{Client selection threhold:}
To facilitate comparison during the simulations, we simplify Eq.~(\ref{eq:final}) by introducing a parameter $\tau$ as the selection threshold. 
The objective is updated to ensure that the cumulative score of selected clients exceeds $\tau_t$, i.e., $ \sum_{k=0}^{K-1} m_{k}^{t} \mathcal{S}(\theta_{k}^{t}) > \tau_t$, while minimizing the number of clients selected. 
In this scenario, $m_{k}^{t}$ is determined by ranking $\mathcal{S}(\theta_{k}^{t})$ values in descending order and selecting clients until the cumulative score exceeds the threshold.
The threshold $\tau$, used in setting the client selection, starts at $0.2$ and increases by $0.1$ every two global rounds.

\noindent\textbf{Performance metric:}
In this paper, our objective is to reduce client participation without compromising model accuracy, setting forth two critical performance metrics: \textit{model accuracy}, which evaluates predictive performance, and \textit{client participation ratio}, measured by the percentage of selected clients relative to all potential participants. This strategy ensures a balance between resource efficiency and learning effectiveness within the FL framework.

\subsection{Performance Analysis Against Baselines}


We evaluate the performance of \textit{FedABC} against the following baseline methods:
1) \textit{FedAvg} \cite{McMahan2017}: a classical method that averages client model parameters;
2) Cho et al. \cite{Cho2022}: a client-selection strategy that prioritizes clients based on the magnitude of their local loss.
The simulation results are presented in Fig.~\ref{fig:com}.

Fig.~\ref{fig:com_part} illustrates that \textit{FedABC} maintains a lower client participation ratio throughout the training process, averaging around 65\% compared to \textit{FedAvg}'s full participation. 
Despite involving fewer clients, \textit{FedABC} achieves comparable or superior server model accuracy across various data heterogeneity, as shown in Fig.~\ref{fig:com_1}, \ref{fig:com_0.5} and \ref{fig:com_0.1}.
As the disparity in data distribution among clients increases (i.e., as $\alpha$ decreases), the performance gap between \textit{FedABC} and the client-selection method based solely on client loss (Cho et al.'s method) becomes more significant.
Specifically, under highly skewed data distributions ($\alpha = 0.1$), \textit{FedABC} outperforms Cho et al.'s method by achieving higher server model accuracy despite involving fewer clients. 
On average, \textit{FedABC} reduces client participation by 2\% compared to Cho et al.'s method while improving accuracy by approximately 3.5\%.
These results demonstrate \textit{FedABC}'s superior performance in enhancing model accuracy while minimizing communication and computation overhead, particularly under high data heterogeneity.7

\subsection{Performance Analysis Under Various Threshold Design}
In this section, we examine how different design strategies for the threshold $\tau$ affect the performance of \textit{FedABC}.
We implement three growth methods (i.e., linear, concave, and convex), while keeping their average client participation ratios approximately equal to ensure a fair comparison.
Specifically, the concave method uses a logarithmic function, while the convex method uses a quadratic function.
Despite similar average participation ratios across all strategies, the convex method results in higher model accuracy across varying degrees of data heterogeneity, as depicted in Fig.~\ref{fig:cifar10}.
This approach starts with a lower client participation ratio and increases it more rapidly, aligning with the ``later-is-better" principle. 

\subsection{Cost Analysis}
\noindent\textbf{Computation cost:}
As depicted in Fig.~\ref{fig:arch}, our approach requires each client to evaluate the server model on its private dataset, adding a computation step beyond standard \textit{FedAvg}.
Model evaluation incurs a low computational cost, as it only requires forward passes without the intensive backpropagation needed for training \cite{Wang2022}.
For the server, our client selection algorithm only requires basic model evaluations and multiplications, easily managed by its ample resources.

\noindent\textbf{Communication cost:}
Each client transmits a small evaluation value (a few bytes) to the server, while the server sends a binary selection indicator (0/1) back to each client, keeping communication load minimal.

In summary, despite the additional steps for computation and communication, the costs introduced by our approach are minimal. 
Compared to the reduced client participation ratio, these costs ensure that the method remains efficient and scalable in practical FL settings.

\section{RELATED WORK}
Client selection is a critical strategy to address data heterogeneity among distributed clients, ensuring that contributions from diverse clients enhance the overall model training process, mitigate biases and improve efficiency \cite{Mayhoub2024}.

Several innovative client selection approaches have been proposed from the view of uneven data distribution.
The AUCTION algorithm \cite{Deng2021} assesses client utility by evaluating local model losses against a global auxiliary dataset, effectively identifying clients whose data are crucial for improving model accuracy. 
Nagalapatti et al. \cite{Nagalapatti2021} select clients based on the  data relevance to specific target labels.
A tier-based method segregates clients based on their training performance, selecting those from similar tiers to address delays caused by resource and data heterogeneity \cite{Chai2020}. 
Additionally, the grouping-based scheduling strategy proposed by Ma et al. \cite{Ma2021} clusters clients so their data labels complement each other, enhancing the diversity and representativeness of the training data.

Furthermore, some methodologies prioritize clients with higher local losses during the model aggregation phase, hypothesizing that these clients' data may lead to more significant improvements in the global model \cite{Cho2022}.
The size of the local datasets is also taken into account, with larger datasets presumed to provide more extensive insights and, therefore, exerting greater influence on the training outcomes \cite{Goetz2019}.

Despite these advancements, many current methods still tend to treat these aspects in isolation and view each global round as an independent event. 
Such a segmented perspective may limit the potential for optimization in FL systems, as it does not fully leverage the continuous learning and adaptability that are inherent to federated settings. 

\section{CONCLUSION}
In this work, we presented \textit{FedABC}, a novel client selection algorithm designed for optimizing FL in heterogeneous and resource-constrained settings.
Our attention-based client selection strategy leverages model similarity and unique contribution to the global model, maximizing learning efficiency while reducing communication and computational costs.
Meanwhile, \textit{FedABC} uses an adaptive client selection threshold, progressively lowering over time based on the ``later-is-better" principle to increase participation in later training stages. 
Extensive simulations demonstrate that \textit{FedABC} achieves high model accuracy with reduced client participation, marking a step forward for practical FL deployment in real-world environments.

\end{document}